# Magnetic Skyrmion Bundles and Their Current-Driven Dynamics

Jin Tang[1], Yaodong Wu[1,2], Weiwei Wang[3], Lingyao Kong[4], Boyao Lv,[1] Wensen Wei[1], Jiadong Zang[5,6], Mingliang Tian[1,4], and Haifeng Du[1,3*]

[1]Anhui Province Key Laboratory of Condensed Matter Physics at Extreme Conditions, High Magnetic Field Laboratory of Chinese Academy of Sciences, and University of Science and Technology of China, Hefei, 230031, China

[2]School of Physics and Materials Engineering, Hefei Normal University, Hefei, 230601, China

[3]Institutes of Physical Science and Information Technology, Anhui University, Hefei 230601, China

[4]School of Physics and Materials Science, Anhui University, Hefei, 230601, China

[5]Department of Physics and Astronomy, University of New Hampshire, Durham, New Hampshire 03824, USA

[6]Materials Science Program, University of New Hampshire, Durham, New Hampshire 03824, USA

*Corresponding author: duhf@hmfl.ac.cn

**Abstract**

Quantization of topological charges determines the various topological spin textures that are expected to play a key role in future spintronic devices. While the magnetic skyrmion with a unit topological charge $Q = \pm 1$ has been extensively studied, spin textures with other integer-valued $Q$ have not been verified well so far. Here, we report the real-space image, creation, and manipulation of a type of multi-$Q$ three-dimensional skyrmionic texture, where a circular spin spiral ties a bunch of skyrmion tubes. We define these objects as skyrmion bundles, and show they have arbitrarily integer values $Q$ from negative up to at least 55 in our experiment. These textures behave as quasiparticles in dynamics for the collective motions driven by electric pulses. Similar to the skyrmion, skyrmion bundles with $Q \neq 0$ exhibit the skyrmion Hall effects with a Hall angle of ~ 62°. Of particular interest, the skyrmion bundle with $Q = 0$ propagates collinearly with respect to the current flow without the skyrmion Hall effect. Our results open a new perspective for possible applications of multi-$Q$ magnetic objects in future spintronic devices.

Magnetic skyrmion is a vortex-like spin texture (Fig. 1a), in which the local spins point in a concerted manner effectively wrapping a sphere[1-3]. The peculiar twists of the spins within the skyrmion imparts an elementary topological charge $Q$, defined as $Q = 1/(4\pi) \int \int d^2\mathbf{r}\, \mathbf{m} \cdot (\frac{\partial \mathbf{m}}{\partial x} \times \frac{\partial \mathbf{m}}{\partial y})$, which essentially counts how many times the magnetization vector field **m** winds around the unit sphere[3]. Its definition in a rotationally symmetric spin texture is clearer if the cylindrical coordinate is introduced. Setting the origin at the rotation center, the spin at location $\mathbf{r} = (r\cos\varphi, r\sin\varphi)$ has the polar angle of $\Theta$ and azimuthal angle of $\Phi$. Due to the rotation symmetry, $\Phi$ has $\varphi$-dependence as $\Phi(\varphi) = v\varphi + \gamma$, where the integer $v$ is the vorticity and the residual angle $\gamma \in [-\pi, \pi]$ is the helicity. On the other hand, $\Theta$ only has $r$-dependence, and the difference $p = \cos\Theta(r = \infty) - \cos\Theta(r = 0)$ define the polarity. The polarity, vorticity, and helicity uniquely define the skyrmion texture and the topological charge is simply $Q = pv$. Such integer-valued topological charge $Q$ induces the Magnus force acting on skyrmions, resulting in a sideway motion dubbed the skyrmion Hall effect[4-6]. The nonzero topological charge of magnetic skyrmion also imparts an emergent electromagnetic field on the electrons passing through it, yielding novel spin-electronic couplings and skyrmion motion under low current density[7]. This property, together with its tunable nanometer-scale size[8], leads to promising applications in future high density and low energy-consumption spintronic devices including skyrmion racetrack memory[9-11], logical devices[12], and artificial synapses for neuromorphic computing[12,13].

Like many elementary particles in nature, a skyrmion only has unit topological charge $Q = \pm 1$[3]. As an elementary quality in topological magnetism, it is a fundamental demand to explore topological spin textures with other integer values of topological charges. Experiments have demonstrated the non-zero high-$Q$ skyrmion cluster states that can be driven by current in the thin plate of chiral magnet FeGe[14]. In addition, there are also theory-predicted metastable multi-$Q$ skyrmionic textures with particle-like behaviors in two-dimensional (2D) chiral magnets[15,16]. A skyrmionium[17-19], or equivalently a $2\pi$-vortex (Fig. 1b), has its spin rotation by an angle of $2\pi$ from the center to periphery. Because of $Q = 0$ for skyrmionium, the skyrmion Hall effect should be absent and the trajectory can be controlled much easily. The high-$Q$ skyrmion bag state, or the skyrmion sack, was theoretically proposed recently[15,16]. Each skyrmion bag comprises with $N$ skyrmions inside a larger circular spin spiral (Fig. 1c). Topological charge of a bag is the summation of that of center skyrmions and the boundary helical stripe, as the integral in its definition applies. Skyrmions inside the bag are conventional, with polarity $p = 1$ and vorticity $v = 1$, so that each skyrmion has the topological charge $Q = 1$. The boundary helical stripe has the same vorticity but opposite polarity, so that $Q = -1$. As a result, the bag has $Q = N - 1$. Owing to their high degrees of freedom and particle-like property, skyrmion bags are exciting new building blocks for spintronics applications such as interconnect devices[16,20-22]. However, despite the fundamental significance and attractive applications of the skyrmion bags[15,16], their experimental realization and dynamical control using electric current remain elusive so far.

Here, we report the unequivocal experimental realization of a type of 3D multi-$Q$

skyrmionics textures and their response to current pulses in the thin plate of B20-type chiral FeGe magnet by using Lorentz transmission electron microscopy (TEM).

In real chiral magnet samples, the third dimension causes profound differences in spin textures. The skyrmion forms tube, which generates monopoles[23], or chiral magnetic bobbers when broken[24], and twisted along the third dimension[25]. In this case, the topological charge $Q$ is specified in each layer of a 3D magnetic configuration because the topological charge is only defined in 2D plane. To explore possibility of multi-$Q$ states and analyze the detailed spin textures therein, simulations based on the measured magnetic parameters of FeGe were performed on a thin FeGe plate with thickness of $t$ = 150 nm (see Method section for the details about the MuMax3 calculations)[26]. Relaxing a uniformly stacked skyrmion bag with a topological charge $Q$ = 5 as the initial state, we obtained a stable 3D texture very different from the simple stacking of skyrmion bags in the thickness direction because of the conical magnetization background (See the supplementary Fig. 1 and note 1). As shown in Fig. 1d, the skyrmion bags only persist in the interior of the sample. While approaching the sample surfaces, the central skyrmions twist and merge with the surrounding spin spiral. Therefore, the skyrmion bag gradually deforms into a complex chiral vortex structure on the sample surface. Detailed analysis shows the surface vortex also has a topological charge of $Q$ = 5 (Supplementary Fig. 2). The topological charge $Q$ actually has the same value in each sliced layer (Shown in Supplementary Fig. 3 and Supplementary Video 1). Given their geometries, these new series of topological textures are termed as the skyrmion bundles, in analogy to vortex bundles in superconductivity[27]. The Lorentz

TEM technique is only sensitive to the variations of in-plane magnetic component integrated along the electron trajectory[28]. The average in-plane magnetization of the simulated $Q = 5$ skyrmion bundle is shown in Fig. 1e, where the magnetic contrast of the boundary spin spiral is weaker than that of the skyrmions inside owing to the aforementioned 3D modulation of the skyrmion bundle.

Considering the simulated results, a thin plate of B20-type FeGe of identical thickness ($t \sim 150$ nm) was employed to demonstrate the formation of skyrmion bundles. FeGe has a high Curie temperature of $T_c \sim 278$ K. The ground state at zero magnetic field is the magnetic spiral with a relatively small period $L_D \sim 70$ nm[25,29]. When a magnetic field **B** is applied perpendicular to the plate, the magnetic spiral states transfer to the magnetic skyrmion tubes of $Q = -1$ ($Q = 1$) for a positive (negative) magnetic field (Supplementary Fig. 4). Since the transition is first-order, the coexisting nonequilibrium state of skyrmions and spin spirals is allowed and is the precursor of skyrmion bundles. Using a field-cooling procedure (Supplementary Fig. 4)[29], we created the mixture state at zero-field and low temperature of $T = 95$ K, as shown in Fig. 2a. Since the negative field was applied for field-cooling, skyrmion tubes with $Q = 1$ (white dots) nucleated from the spiral background. Afterwards, the positive field was gradually turned on. Although $Q = 1$ skyrmion tubes are not thermodynamically stable at the positive fields, they can typically survive in the nonequilibrium metastable state in moderate positive fields (Fig. 2b). With a further increase of $B \sim 200$ mT, most skyrmion tubes disappear but the ones left, which are represented by white dots and form a cluster of six skyrmions encircled by a dark ring (Fig. 2c). Using the transport of intensity equation (TIE) analysis[28], we obtained its in-plane magnetization mapping, where the surrounding circular spiral with weak contrast (Fig. 2d) has opposite rotation sense compared to the skyrmions inside. The experimental result is in good agreement with the numerical one (Fig. 1e), clearly indicating a skyrmion bundles of $Q = 5$. The skyrmion bundle persists after its creation even when the field is reduced. Repeating all over from the field-cooling, skyrmion bundles can be always generated in this manner, but the number of skyrmion tubes in the bundle will vary and depend on the initial magnetic configurations.

The topological charge of the skyrmion bundle can be tunable using cascading topological transitions enabled by a magnetic field. A skyrmion bundle with 18 skyrmions inside was used for a representative demonstration (Fig. 2e). It persists up to $B \sim 160$ mT. Further increment of the field annihilates one skyrmion tube inside, while the boundary spin spiral remains robust. The topological charge is thus reduced by 1. Such topological transitions of successively killing skyrmions continue in a broad field range of 160 mT $< B <$ 320 mT, until a magnetic bundle with $Q = 0$ appears at $B \sim 235$ mT. Such skyrmion bundle has only one skyrmion tube wrapped inside it (Supplementary Fig. 3), therefore, it is the generalization of the skyrmionium in 3D. Interestingly, upon further increasing the field to ~321 mT, the internal skyrmion tube disappears and the outside spin spiral shrinks to a normal skyrmion tube with $Q = -1$. This behavior validates the topological equivalence between the surrounding spin spiral and the $Q = -1$ skyrmion in spite of their visible difference in magnetic configurations.

Fig. 2f shows the field-driven evolution of the topological charge. Reversing the fields in the entire process can generate skyrmion bundles with negative topological charges (Supplementary Fig. 5b). By combing the field-cooling process and the field-driven cascading topological transition, a complete set of topological charge values was covered with bundles containing up to 56 skyrmions. (Fig. 2g and Supplementary Fig. 5).

The multi-$Q$ skyrmion bundles discovered in this study are not only novel magnetostatic states but also quasiparticles in current-driven dynamics. A microdevice composed of a 150 nm thick FeGe thin plate and two electrodes was fabricated for the *in-situ* Lorentz TEM observation (Supplementary Fig. 6). All skyrmion bundles moved collectively under the current.

Fig. 3a shows the representative sequence of selected Lorentz-TEM images of a $Q = 18$ skyrmion bundle after successive nanosecond pulsed stimulations with a typical current density $j \sim 4.0 \times 10^{10}$ A m$^{-2}$ applied in the $-x$ direction. The pulse width is set to be 80 ns with a frequency of 1 Hz (Supplementary Fig. 6). Despite a slight rotation, the skyrmion bundle moved as a rigid body in the $x$ direction, opposite to the direction of the current flow (Supplementary Videos 2). It moved in the opposite direction when the current was reversed (Fig. 3b and Supplementary Video 3). A transverse motion in analogy to the skyrmion Hall effect was also observed[4-6]. Both longitudinal motion and the Hall effect were observed up to $Q = 55$ bundle in our experiment (Supplementary Video 4). The dynamics of single magnetic bundle are similar to the skyrmion aggregate reported by Yu *et al.*[14]. However, they are two different magnetic objects in essence. Magnetic bundles are stable magnetic objects with particle-like dynamical behaviors, and keep separable and stable even when they get close to each other. In contrast, skyrmion aggregates easily merge when they get close owing to the attractive forces between each other (Supplementary Fig. 7 and Video 5) [25], giving rise to the stochastic nature of the skyrmion number. Moreover, the aforementioned field-induced cascading topological transition and the skyrmionium with $Q = 0$ are far beyond the physics of simple assembly of normal skyrmions.

Trajectory of the $Q = 18$ skyrmion bundle reveals a linear relation between the displacement and time (Fig. 3c), as expected for normal skyrmions[5,6]. Linear fitting under various current density $j$ showed that both the longitudinal and transverse velocities, $v_x$ and $v_y$, are proportional to the current density $j$ (Fig. 3d). The skyrmion Hall effect is quantitatively described by the Hall angle $\theta_h$ defined as $\tan\theta_h = v_y/v_x$[5]. The $\theta_h - j$ curve indicates that $\theta_h$ is about $60°$ (Fig. 3e).

The threshold current density $j_{c1}$ required for the motion of skyrmion bundles is about $\sim 2.5 - 4.0 \times 10^{10}$ A m$^{-2}$ depending on the topological charge $Q$. Small objects including skyrmionium, single skyrmion and low-$Q$ skyrmion bundles have in general high values, suggesting that they are more easily pinned by localized defects (Supplementary Fig. 8). Such threshold current density is two orders of magnitude smaller than that required for domain walls[30], and even one order of magnitude smaller than that of interfacial skyrmions[31-35]. When $j > j_{c2} = 4.5 \times 10^{10}$ A m$^{-2}$, skyrmion bundles become unstable during their motion, which can be attributed to inevitable Joule heating at large currents[36].

The longitudinal motions of both the $Q = 0$ skyrmion bundle, or equivalently 3D

skyrmionium and the single skyrmion were observed as well, as shown in Figs. 4a and 4b. Excitingly, the skyrmion Hall effect does vanish for skyrmionium owing to its zero $Q$ and vanishing Magnus force (Supplementary Video 6)[17,18]. The slight deviation of the trajectory from the *x* direction can be reasonably attributed to the pinning of defects or disorders. More importantly, under reversed current orientation, collinear motion in the opposite direction was also observed (Supplementary Fig. 9), further supporting the absence of the skyrmion Hall effect. The single skyrmion tube nucleated from the 3D skyrmionium has $Q$ = -1, and hence, the skyrmion Hall effect reappears but the transverse motion is in the **-y** direction instead (Fig. 4b and Supplementary Video 7), which is consistent with the theoretical prediction[4].

The trajectories of the magnetic skyrmion bundles ($Q > 0$), 3D skyrmionium ($Q = 0$), and skyrmion ($Q = -1$) are summarized in Fig. 4c. The skyrmion Hall effect is clearly seen for $Q \neq 0$ skyrmionic textures and the vanishing skyrmion Hall effect for the $Q = 0$ 3D skyrmionium. The $Q$-dependent Hall angle is shown in Fig. 4d, where the sign of Hall angle is determined by the sign of the topological charge. Moreover, the absolute value of Hall angle $|\theta_h| = 62° \pm 5°$ is almost independent of current density $j$, external field $B$ (Supplementary Fig. 10), and nonzero $Q$ (Fig. 4d and Supplementary Fig. 11). Such universal behavior of the skyrmion Hall effect in bulk materials is completely different from the previous reports on interfacial skyrmions or skyrmion bubbles in magnetic multilayers, wherein Hall angle $\theta_h$ depends on current density $j$ and external field $B$[5,6,33,35]. For skyrmion bubbles in multilayers with diameters $d$ and domain wall widths $\gamma_{dw}$ ($\gamma_{dw} < d$), the shape factor $\eta$ defined as $\eta = \pi^2 d/(8\gamma_{dw})$ may be strongly affected by the deformations of moving skyrmion bubbles[5,6,33,35], resulting in a drive-dependent Hall effect.

For the multi-$Q$ skyrmion bundles, the Hall angle in the real system is the sum of intrinsic and extrinsic contributions (See the details in the Supplementary Note 2)[37,38]. The intrinsic $\theta_h$ is for the system without impurities and can be approximately written as $\theta_h \approx \arctan\left[(\alpha - \beta)\frac{\eta}{Q}\right]$ with the intrinsic material parameters of Gilbert damping $\alpha$ and nonadiabatic constant $\beta$, respectively[20]. Numerical calculations show that $\eta$ of bundles is approximately proportional to $Q$ (Fig. 5a), so that $\theta_h$ is not strongly dependent on topological charge $Q$. 3D dynamical simulations using measured magnetic parameters confirm the weak $Q$-dependence of $\theta_h$. Derivation of $\theta_h$ depends on magnetic parameters and is reduced to ~2° as the coefficient $(\alpha - \beta)$ increases to 0.65 (Fig. 5b). Such a relatively small variation of $\theta_h$ is consistent with the results of 2D skyrmion bags[20].

Such small variation of $\theta_h$ can be hardly addressed in the present experiments because impurities in real materials could induce the extrinsic Hall angle[38], which can be roughly estimated to be around (50° ~ 65°). Generally, the value of α is typically smaller than 0.1 in ferromagnetic metals[38-40], so that the intrinsic Hall angle is $\theta_h < 15°$. So the total $\theta_h$ is roughly in agreement with the experimental value of $62 \pm 5°$. Note that the quantitative calculation of $\theta_h$ in the real system is difficult and beyond the scope of present work owing to the unknown $\alpha - \beta$ and the lack of exact information of impurities.

We expect that skyrmion bundles can be realized in a wide range of chiral magnets[41]. Variable multi-$Q$ magnetic states tunable by the magnetic field and drivable by the electric current hold great promise for future spintronic devices, such as skyrmionium-based racetrack memory and atomic-nucleus-like information encoding[16,18]. In addition to the collective motions as quasiparticles, further advance in the electrical reading and writing of the zoo of multi-$Q$ skyrmion bundles invites extensive future investigations.

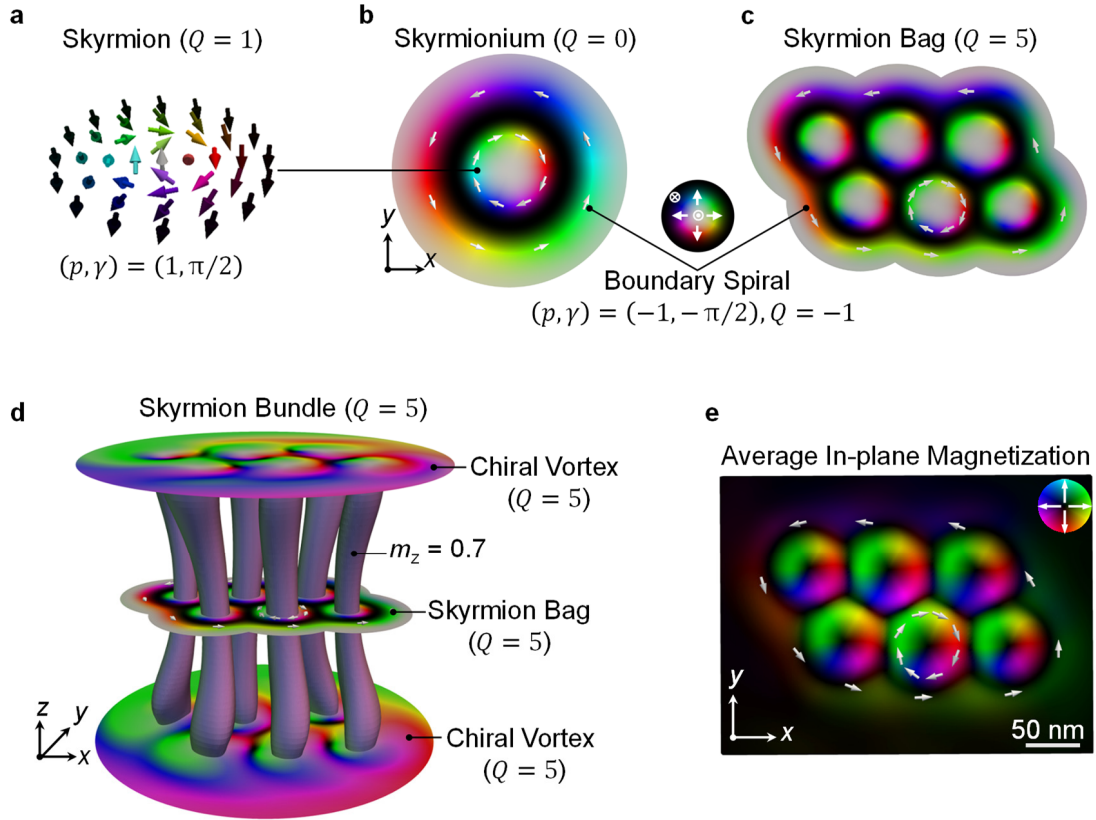

**Fig. 1 | Schematics of a magnetic skyrmion bundle. a-c**, Representative two-dimensional skyrmionic textures of a skyrmion with $Q = 1$ (**a**), skyrmionium with $Q = 0$ (**b**), skyrmion bag with $Q = 5$ (**c**). **d**, Schematic of a $Q = 5$ skyrmion bundle in a thin FeGe plate with the thickness of 150 nm. Chiral magnetic vortex structures appear around the two surfaces, and the interior are the skyrmion bags. The internal skyrmion tubes are shown by the iso-surfaces of normal magnetization $m_z = 0.7$. **e**, The average in-plane magnetization mapping over the depth of the skyrmion bundle that can be obtained by Lorentz-TEM measurements. The colors in **a–d** represent the magnetization orientation according to the color wheel in **b**. Dark and white contrasts stand for the magnetizations are down and up, respectively. The colors in **e** represent the in-plane magnetizations according to the color wheel in **e**; dark contrasts stand for the in-plane magnetizations are zero. The scale bar in **e** is 50 nm.

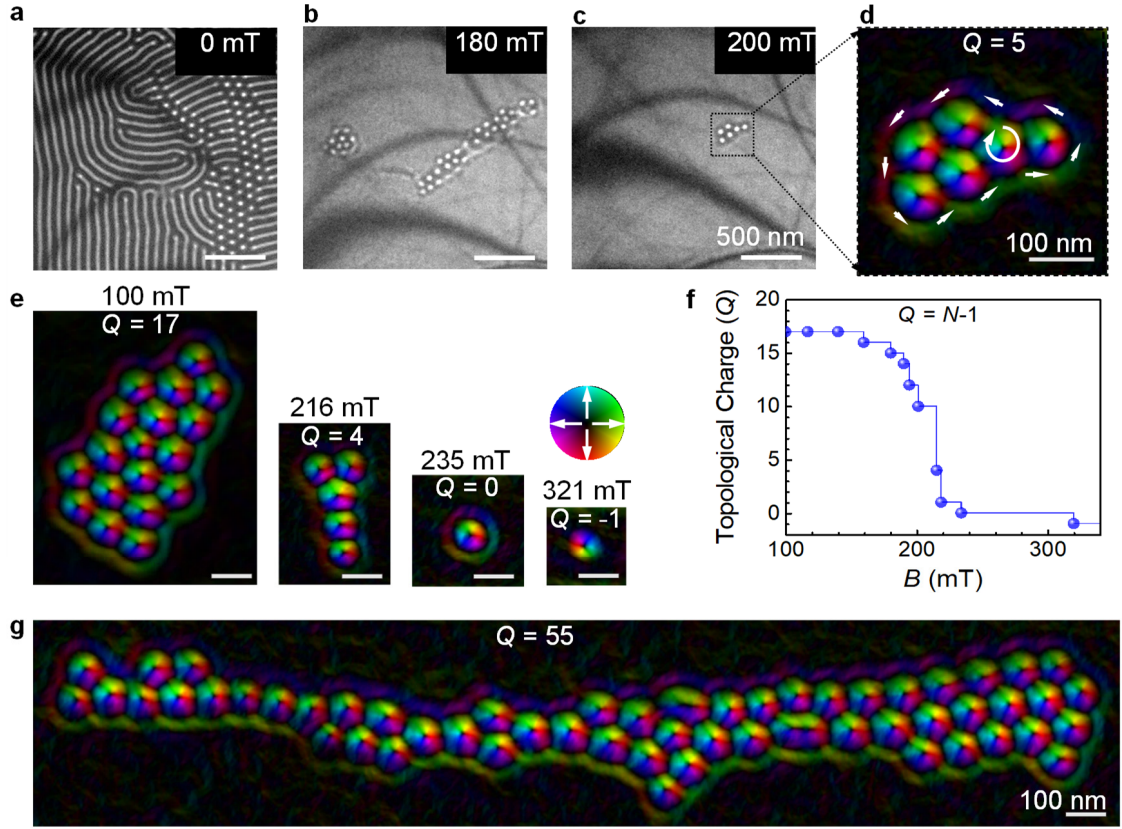

**Fig. 2 | Realization of magnetic skyrmion bundles with varying topological charges.**

**a**, Lorentz TEM images of mixed skyrmion tubes with $Q = 1$ (marked by white dots) and spin spirals in a thin FeGe plate with a thickness of $t \sim 150$ nm at zero field and temperature $T = 95$ K realized by a negative field cooling process. **b**, A complex cluster of skyrmion tubes at $B \sim 180$ mT. **c**, A complete magnetic skyrmion bundle with $Q = 5$ at $B \sim 200$ mT. **d**, Magnified in-plane magnetization mapping of the magnetic skyrmion bundle with $Q = 5$ in the marked zone of **c**. **e**, Typical magnetic configurations of field-driven cascading quantized transitions from a skyrmion bundle with $Q = 17$ to a single skyrmion with $Q = -1$. A bundle with $Q = 0$ forms at $B \sim 235$ mT, and this is the 3D skyrmionium. The single skyrmion is the limiting case of skyrmion bundle without the central skyrmion tube. **f**, Topological charge $Q$ as a function of magnetic field $B$ applied perpendicular to the plate plane. $N$ represents the number of interior skyrmion tubes. **g**,

A representative magnetic skyrmion bundle with a topological charge of $Q = 55$ at $B \sim 100$ mT. Lorentz images in **a-c** were obtained using the out-of-focus condition with the defocus value of -1000 μm. The magnetic configurations in **d**, **e**, and **g** are represented by in-plane magnetization mappings obtained by TIE analysis from the Lorentz TEM images. The color wheel in **e** represents in-plane magnetization distributions. The scale bars in **a-c** and **d, e, g** are 500 nm and 100 nm, respectively.

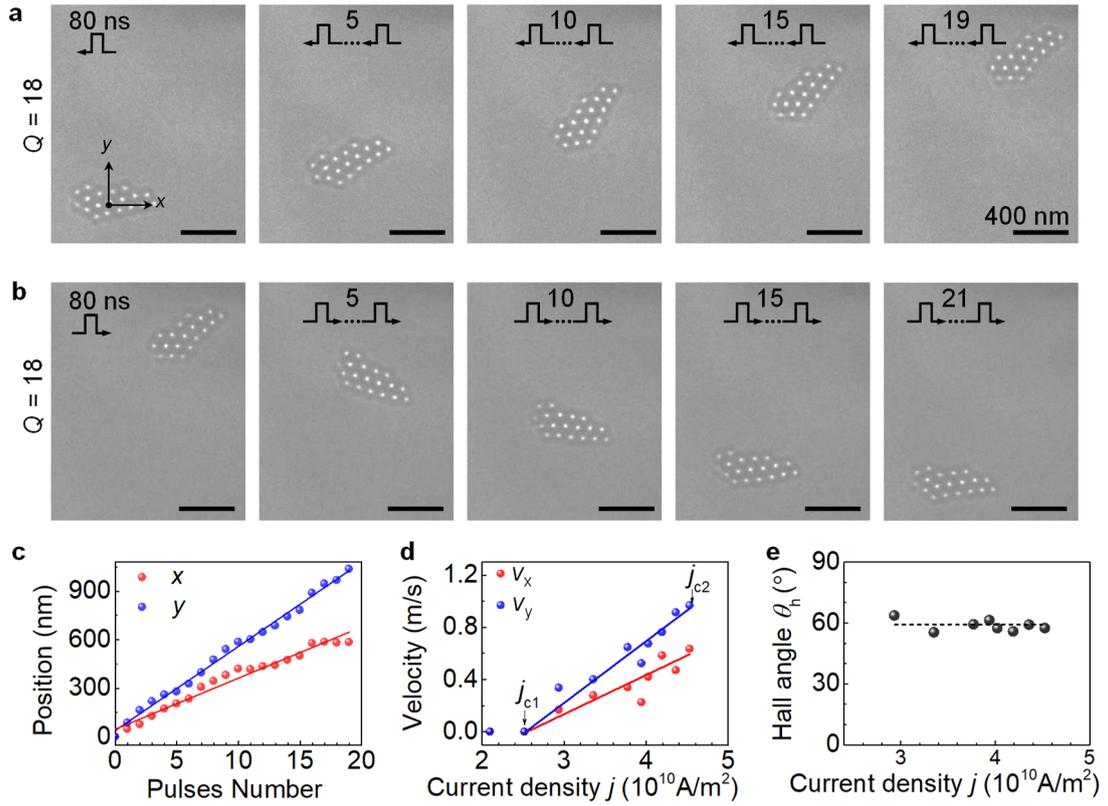

**Fig. 3 | Current-driven motions of a skyrmion bundle with $Q$ = 18. a**, Snapshots of the current-driven motions of a skyrmion bundle at a current density of $j \sim 4.0 \times 10^{10}$ A m$^{-2}$ along the -$x$ axis and at $B \sim 100$ mT. The bundle moves along the $x$ direction, accompanied by a transverse motion along the $y$ direction. **b**. Snapshots of current-driven motions of the skyrmion bundle with reversed current. **c**. Dependence of positions $x$ and $y$ of the skyrmion bundle on the pulse number. **d**, Dependence of average velocities along the $x$ ($v_x$) and $y$ ($v_y$) axes on current density $j$. **e**, Skyrmion Hall angle $\theta_h$ as a function of current density. Lorentz images in **a, b** were obtained under the out-of-focus conditions with a defocus value of -1000 μm. The number of current pulses is marked at the top of the corresponding panels. The scale bars in **a** and **b** are 400 nm.

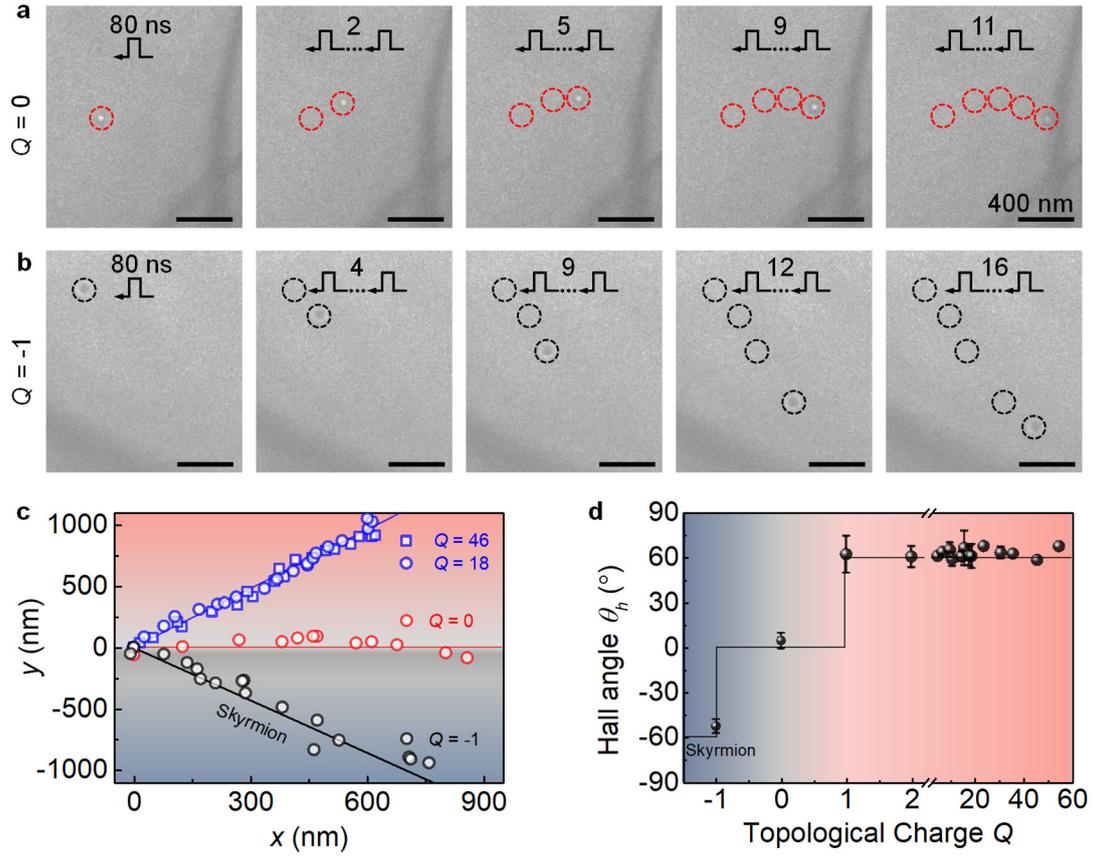

**Fig. 4 | Dependence of the skyrmion bundles dynamics on the topological charge Q. a, b**, Snapshots of current-driven motion of a 3D skyrmionium or $Q = 0$ skyrmion bundle (**a**), and a single $Q = -1$ skyrmion (**b**) at the current density $j \sim 4.2 \times 10^{10}$ A m$^{-2}$ along the -*x* direction and $B \sim 100$ mT. The number of current pulses is marked in the top of the corresponding panels. **c**, Representative trajectories of skyrmion bundles ($Q > 0$, blue dots), 3D skyrmionium ($Q = 0$, red dots), and skyrmion ($Q = -1$, black dots). **d**, Dependence of the Hall angle $\theta_h$ on the topological charge $Q$. The error bars in **d** correspond to the standard deviation obtained from multiple measurements. The previous positions and trajectories are marked by circles in **a** and **b** to guide the eyes. The scale bars in **a** and **b** are 400 nm.

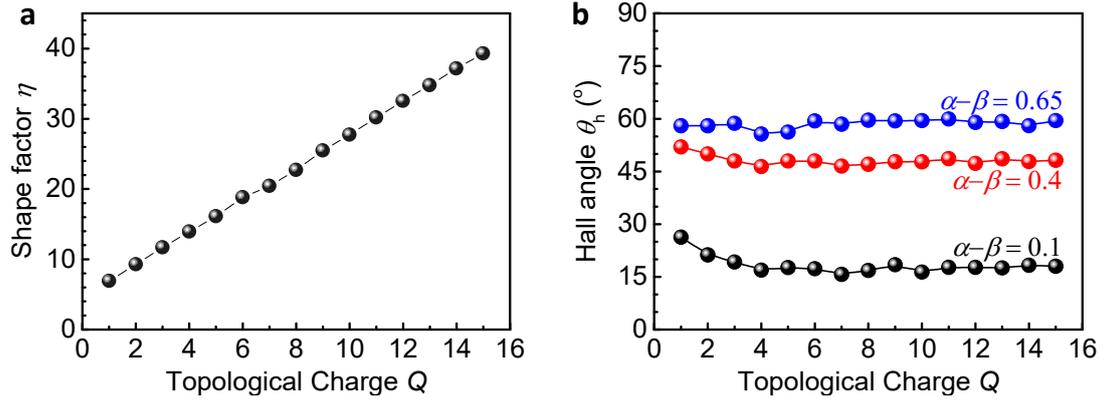

**Fig. 5 | Simulated skyrmion Hall effects of skyrmion bundles. a**, Dependence of the averaged shape factor $\eta$ on the topological charge $Q$. **b**, Dependence of Hall angle on topological charge using STT effect for $(\alpha - \beta) = $ 0.1, 0.4 and 0.65, respectively.

## Methods
### Preparation of FeGe crystals

B20 type FeGe single crystals were grown by chemical vapor transport method with mixture of stoichiometric iron (Alfa Aesar, purity > 99.9%), germanium (Alfa Aesar, purity > 99.99%) and transport agent $I_2$. The B20 type FeGe crystallize in a temperature gradient from 560 °C to 500 °C, with a nearly pyramidal shape for space group $P2_13$.

### Fabrication of FeGe micro-devices

Thin FeGe thin plates of thickness of $t$ ~150 nm for TEM magnetic imaging were fabricated by the lift-out method using a focused-ion beam and scanning electron microscopy dual beam system (Helios Nanolab 600i, FEI) in combination with a gas injection system and a micromanipulator (Omniprobe 200+, Oxford).

### TEM measurements

Magnetic imaging was carried out using a TEM instrument (Talos F200X, FEI) operated at 200 kV in the Lorentz Fresnel mode. The objective lens is switched off to provide a field-free condition. A single-tilt liquid-nitrogen specimen holder (Model 616.6 cryotransfer holder, Gatan) was used with a temperature range from 95 to 300 K. The current pulses were provided by a voltage source (AVR-E3-B-PN-AC22, Avtech Electrosystems Ltd.) and were set to be 80 ns with a frequency of 1 Hz. The experiments described in the main text were all performed at the temperature of 95 K.

### Micromagnetic simulations[26]

Micromagnetic simulations were performed using the Mumax3. The total free energy terms are written as:

$$\varepsilon = \int_{V_s} \{\varepsilon_{\text{ex}} + \varepsilon_{\text{DMI}} + \varepsilon_{\text{zeeman}} + \varepsilon_{\text{dem}}\} d\boldsymbol{r} \tag{1}$$

Here, exchange energy $\varepsilon_{\text{ex}} = A(\partial_x \mathbf{m}^2 + \partial_y \mathbf{m}^2 + \partial_z \mathbf{m}^2)$, DMI energy $\varepsilon_{\text{DMI}} = D_{\text{dmi}} \mathbf{m} \cdot [\nabla \times \mathbf{m}]$, Zeeman energy $\varepsilon_{\text{zeeman}} = -M_s \mathbf{B}_{\text{ext}} \mathbf{m}$, and demagnetization energy $\varepsilon_{\text{dem}} = -\frac{1}{2} M_s \mathbf{B}_d \mathbf{m}$. Here $\mathbf{m} \equiv \mathbf{m}(x, y, z)$ is the normalized units continuous vector field that represents the magnetization $\mathbf{M} \equiv M_s \mathbf{m}(x, y, z)$. $A$, $D_{\text{dmi}}$, and $M_s$ are the exchange interaction, DMI interaction, and saturation magnetization, respectively. $\mathbf{B}_d$ is the demagnetizing field. We set a typical value for saturation magnetization $M_s$ = 384 kA m$^{-1}$ for FeGe[25]. The exchange interaction $A_{\text{ex}}$ = 3.25 pJ/m is determined from the fit to field-dependence of magnetization evolution[25]. DMI interaction $D_{\text{dmi}} = 4\pi A/L_D$ = 5.834 mJ m$^{-2}$ is obtained from zero-field spin spiral period $L_D$ = 70 nm[25]. We set the cell size as 4 × 4 × 3 nm$^3$. We obtained the equilibrium spin configurations using the conjugate-gradient method.

**Data availability**

The data that support the plots provided in this paper and other findings of this study are available from the corresponding author upon reasonable request.

**Acknowledgments**

H. D. acknowledges the financial support from the National Key R&D Program of China, Grant No. 2017YFA0303201; the Key Research Program of Frontier Sciences, CAS, Grant No. QYZDB-SSW-SLH009; the Key Research Program of the Chinese Academy of Sciences, Grant No. KJZD-SW-M01; the Strategic Priority Research Program of Chinese Academy of Sciences, Grant No. XDB33030100; and the Equipment Development Project of Chinese Academy of Sciences, Grant No. YJKYYQ20180012. J. T. and L. K. acknowledge the financial support of the Natural Science Foundation of China, Grant Nos. 11804343 and 11974021. A portion of this work was supported by the High Magnetic Field Laboratory of Anhui Province.

**Author contributions**

H.D. supervised the project. H.D. and J.T. conceived the experiments. W-S.W. synthesized FeGe single crystals. J.T. and Y.W. fabricated FeGe micro-devices and performed TEM measurements. J.T. performed simulations. H.D., J.T., and J.Z. prepared the manuscript. All authors discussed the results and contributed to the manuscript.

**Competing interests**

The authors declare no competing interests.

**Additional information**

Supplementary information is available for this paper.